\documentclass[pre,twocolumn,aps,showpacs,floatfix,nofootinbib,
superscriptaddress]{revtex4-2}
\usepackage{subfigure}
\usepackage{amssymb}
\usepackage{amsfonts}
\usepackage{amsmath}
\usepackage{amsthm}
\usepackage{epsfig}
\usepackage{graphicx}

\usepackage[usenames,dvipsnames]{color}
\usepackage[utf8]{inputenc}

\usepackage[hidelinks,unicode=true]{hyperref}
\hypersetup{colorlinks=true,% Don't draw a box around links, instead color every piece of text, which is a link
 	linkcolor=blue,
 	urlcolor=blue,
 	citecolor=blue,
 	pdfhighlight=/N% When clicking a link and holding the mouse button, don't use any graphical effects - i.e. the alternative would be to behave like a "button" which is pressed within the PDF
 }
 \usepackage{comment}
 \usepackage[normalem]{ulem}

\newcommand{\lrangle}[1]{\langle{#1}\rangle}

\begin{document}

\title{Bifractality in one-dimensional Wolf-Villain model}

\author{Edwin E. Mozo Luis}
\email{emozo@id.uff.br}
\address{Instituto de F\'{\i}sica, Universidade Federal Fluminense,
  Avenida Litor\^{a}nea s/n, 24210-340, Niter\'{o}i, RJ, Brazil}

 \author{Silvio C. Ferreira}
 \email{silviojr@ufv.br}
 \affiliation{Departamento de F\'{\i}sica, Universidade Federal de
 	Vi\c{c}osa, Minas Gerais, 36570-900, Vi\c{c}osa, Brazil}
 \affiliation{National Institute of Science and Technology for Complex Systems, 22290-180, Rio de Janeiro, Brazil}

 \author{Thiago A. de Assis}
\email{thiagoaa@ufba.br}
\address{Instituto de F\'{\i}sica, Universidade Federal da Bahia,
  Campus Universit\'{a}rio da Federa\c c\~ao,
  Rua Bar\~{a}o de Jeremoabo s/n, 40170-115, Salvador, BA, Brazil}

   \address{Instituto de F\'{\i}sica, Universidade Federal Fluminense,
  Avenida Litor\^{a}nea s/n, 24210-340, Niter\'{o}i, RJ, Brazil}

\begin{abstract}
We introduce a multifractal optimal detrended fluctuation analysis to study the scaling properties of the one-dimensional Wolf-Villain (WV) model for surface growth. This model produces mounded surface morphologies for long time scales (up to $10^9$ monolayers) and its universality class remains controversial. Our results for the multifractal exponent $\tau(q)$ reveal an effective local roughness exponent consistent with a transient given by the molecular beam epitaxy (MBE) growth regime and Edward-Wilkinson (EW) universality class for negative and positive $q$-values, respectively. Therefore, although the results corroborate that long-wavelength fluctuations belong to the EW class in the hydrodynamic limit, as conjectured in the recent literature, a bifractal signature of the WV model with an MBE regime at short wavelengths was observed.

\end{abstract}

\maketitle

\textit{Introduction}. Various surface morphologies can be observed in film growth processes under nonequilibrium conditions~\cite{Krug97,Ohring2002,Evans2006}. Understanding the microscopic mechanisms behind these morphologies is crucial, and lattice models have proven to be valuable tools for elucidating their formation. Additionally, they can be described, in the continuous limit, by stochastic equations~\cite{Krug97, Spohn2015, Oliveira2019, Mallio2022}. Nevertheless, these lattice models may exhibit long transients due to, for example, the presence of mounded surface topographies of the deposit. Hence, their universality classes, formally found in the hydrodynamic limit of large scales and long times, are not easily achieved from the scaling of statistical quantities of interest such as global or local roughness~\cite{Chame2004,PRE2023}. A clear example whose universality class is challenging to access is the Wolf-Villain (WV) model \cite{Wolf_90}. It was proposed to mimic surface adatom diffusion during film growth for low deposit temperatures and has been studied for decades ~\cite{Schroeder93,Pavel94,Kotrla96,Huang96,Punyindu98,Punyindu02,Haselwandter07,Haselwandter08,Ferreira2019,xun12}. The WV model favors the aggregation of adatoms at points on the surface with higher coordination (local minimization of energy).

The squared global surface roughness of the deposit is defined by $W^{2}(L,t) \equiv \langle \overline{h^2(x,t)} - \overline{h(x,t)}^2\rangle$, where $ h(x,t)$ is the height of a column of a deposit at the position $x$ and time $t$, $L$ is the lateral size of the substrate of dimension $d$, and overbars and brackets indicate the spatial and configurational averages, respectively. In the hydrodynamic limit, one expects $W^{2} \sim t^{2\beta}$, where $\beta$ is the growth exponent. At sufficiently long times, $W(L,t)$ is expected to reach a steady state with saturated roughness that scales as $W_s \sim L^{\alpha}$, where $\alpha$ is the global roughness exponent. The crossover between growth and saturation regimes is separated by a characteristic time that scales as $\tau\sim L^z$, where $z=\alpha/\beta$ is the dynamic exponent given by  the Family-Vicsek ansatz  \cite{Family85,Barabasi,Reis13}.

Surface fluctuations can be analyzed locally using the local squared surface roughness, defined by $\omega^{2}(r,t)= { \left\langle \left\langle \left[  h(x,t) - \overline{h}_{r} \right]^{2}   \right\rangle_{r} \right\rangle }$ ~\cite{Juan97,Chame2004}, where $\overline{h}_{r}$ is the average height within a window of size $r$ and  $\left\langle .\right\rangle_{r} $ is the average over different windows of size $r$. If Family-Vicsek scaling is observed then
\begin{equation}
\omega(r,t)\sim t^{\beta}f\left( \frac{r}{t^{1/z}}\right) \sim
\begin{cases}
 t^\beta, &\text{ for~~ } t \ll r^{z}\\
r^{\alpha_{l}}, &\text{ for~~ } t \gg r^{z},\\
\end{cases}
\label{Eq:Fvl}
\end{equation}
with $\alpha_{l}=\alpha$, where $\alpha_{l}$ is the local roughness exponent. Otherwise, an anomalous scaling regime with $\alpha\ne\alpha_{l}$ is at work~\cite{Juan99,Juan05}.

The WV model was proposed initially as a model for molecular beam epitaxy (MBE) where the leading mechanism is diffusion~\cite{Wolf_90}. P\v{r}edota and Kotrla~\cite{Kotrla96} derived analytically, using regularization procedure for $d=1$, the corresponding Langevin equation for the WV model, identifying that it can belong to the Edwards-Wilkinson (EW) universality class~\cite{Edwards82} with exponents $\alpha=1/2$, $\beta=1/4$, and $z=2$. A hydrodynamic limit and its crossovers can addressed using phenomenological Langevin equations. A general equation takes the form~\cite{Barabasi}
\begin{equation}
\frac{\partial h}{\partial t} =  -K\nabla^4 h+\tilde{\lambda}\nabla^2(\nabla h)^2+\nu \nabla^2h+\eta,
\label{eq:sto_eq}
\end{equation}
where the first and second terms are linear and nonlinear contributions due to surface diffusion, $\nabla^2$ is due to surface tension and $\eta$ is an uncorrelated white noise.
For $K=\tilde{\lambda}=0$, Eq.~\eqref{eq:sto_eq} becomes the EW equation~\cite{Edwards82}, for $\nu=\tilde{\lambda}=0$ the Mullis-Hering (MH) equation~\cite{MH2004}, finally, for $\nu=0$ the Villan-Lais-Das Sarma (VLDS) equation~\cite{Villain91,Lai91}. Scaling analysis implies that the surface tension is the leading term in hydrodynamic limit, followed by nonlinear and linear diffusion terms in this sequence~\cite{Barabasi}. However, for small surface tension, an MBE regime is observable for long times and the EW scaling emerges only for extremely long times. Indeed, Costa~\textit{et al.}~\cite{Costa2003} observed a significant reduction in the effective global roughness and dynamic exponents in $d = 1$, indicating asymptotic values consistent with the EW class. Crossovers from the MH~\cite{MH2004} to the VLDS~\cite{Villain91,Lai91} models, reaching the EW class~\cite{Edwards82}, were also reported \cite{Haselwandter07,Haselwandter08}.

An intrinsic anomalous scaling has been considered in the WV model in $d=1$ with $\alpha_l\ne \alpha$~\cite{Xun2010}. However, Xun \textit{et al.} \cite{xun12} studied the WV model in $d=1$, and claimed an asymptotic dynamic scaling given by $\alpha = 0.50(2)$ and $\beta = 0.25(2)$, consistent with the one-dimensional EW equation, considering sizes $L \geqslant 2048$ and times up to than $t \approx 10^9$ monolayers. Nonetheless, in the determination of a local roughness exponent, they used the height difference correlation function, assuming a scaling in distances of the same order as the correlation length of the interface, $\xi$. However, the height difference correlation function scaling is strictly applicable for distances $r\ll \xi$  in the hydrodynamic limit~\cite{Zhao2001,Mallio2022,Mozo2022,PRE2023} [see Fig. S1 in the Supplementary Material (SM)]. Therefore, the WV model scaling remains incompletely understood and exhibits a complex behavior at the time scales reported in the literature. For example, adding a perturbation to the WV rules where surface diffusion in the direction normal to the substrate is allowed, destroy the self-affine properties of the surface, leading to a mounded morphology~\cite{Ferreira2019}.

Motivated by the aforementioned discussion and considering that the WV model exhibits a mounded-type morphology, we introduce the multifractal optimal detrended fluctuation analysis (MF-ODFA) to investigate the scaling properties for both small and large wavelength fluctuations. It is an extension of optimal detrended fluctuation analysis (ODFA), recently introduced in Ref. ~\cite{Luis17}, that enables the extraction of the local roughness exponents at spatial scale lengths $r \lesssim  \xi$ and was successfully applied to investigate lattice models in the VLDS class in $d=1$ ~\cite{Luis19} and $d=2$ ~\cite{Luis17}. Furthermore, the ODFA method does not alter the scaling for length scales $r \gtrsim  \xi$,  preserving global exponents ~\cite{Luis19, Luis17}.

\begin{figure}
    \includegraphics[width=6cm, height=2.8cm]{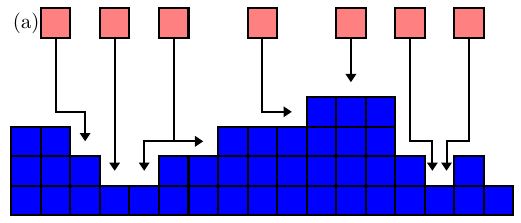} \\ 
    \includegraphics[width=0.65\linewidth]{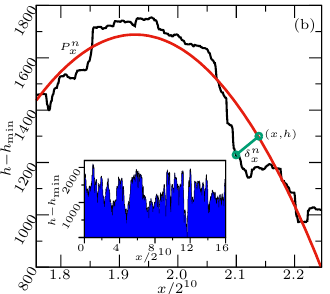}
    \caption{(color online) (a) Deposition rules for the WV model in $d=1$. (b) Schematic representation of the ODFA method. The solid (red) line represents the regression used to detrend an interval of the interface depicted in black. Inset displays the (shifted) whole profile for the WV model at $t=10^{8}$.}
	\label{fig:models_prof}
\end{figure}

\textit{Methods}. The model is defined on a one-dimensional lattice with parameter $a$, with a constant flux of atoms $F$  in the direction perpendicular to the substrate. All spatial quantities discussed here are given in units of $a$. In the WV model, a site of the substrate is randomly chosen and, subsequently, a particle is irreversibly attached at the top of a column chosen among the incident site and its nearest neighbors that leads to the largest number of lateral neighbors.  In the case of a tie between the neighbors, one of them is selected randomly. Time unity is commonly given in units of $F^{-1}$, i.e., in units of deposited monolayers. In our simulations, periodic boundary conditions are considered. The rules for this model are illustrated in Fig. \ref{fig:models_prof}(a) for $d=1$, as considered in this work.

A typical interface with a mounded envelope generated for the WV model at $t=10^8$ is shown in Fig.~\ref{fig:models_prof}(b). The mounded morphology is confirmed with the autocorrelation function, which is defined as $\Gamma(r,t) = \lrangle{\tilde{h}(x+r,t)\tilde{h}(x,t)}$, where $\tilde{h} \equiv h - \bar{h}$ and $\bar{h}$ is the mean height of the profile that presents oscillations as a function of distance [see Fig. S2 in the SM]. Of course, $\Gamma (0,t) = W^2(L,t)$. A characteristic mound width, $\xi_0(t)$, can be defined as the position of the first zero of $\Gamma(r,t)/\Gamma(0)$ [$\Gamma(0) \equiv \Gamma(0,t)$]~\cite{Murty2003, mound, Leal2011, Luis17, Luis19}. 

\begin{figure*}[!t]
	\center
	\includegraphics [width=0.3185\linewidth]{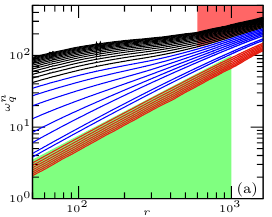} ~ ~ 
	\includegraphics [width=0.3185\linewidth]{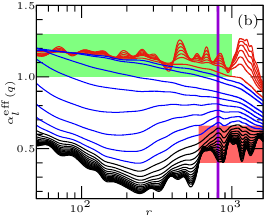} ~ ~ 
	\includegraphics [width=0.28\linewidth]{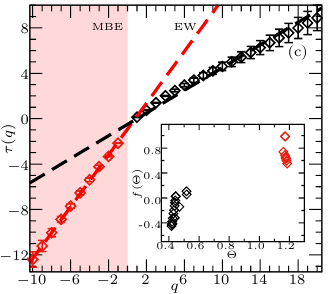}
	\caption{(color online) (a) Local roughness analysis for one-dimensional  WV model with $t=10^8$, $L=2^{14}$,  $-10\leq q \leq 20$ increasing from bottom to top curves, through MF-ODFA method with $n=2$. Averages were computed over $2000$ independent realizations. (b) Local slope analysis of $\alpha_\mathrm{l}^{\mathrm{eff}}(q)$. The vertical line denotes the value of the characteristic mound width $\xi_0(t=10^8) = 808$; see Fig. S2 of the SM. Red and black lines correspond to $-10 \leq q \leq -4$ and $6 \leq q \leq 20$, respectively. The shaded regions indicate the scaling range from which $\alpha_l(q)$ were extracted. Blue curves ($-3 \leq q \leq 5$ with $q \neq 0$) do not present scaling. (c) Multifractal exponent $\tau(q)$ for MF-ODFA analysis of the WV model. Error bars are smaller than symbols. The slopes of the dashed lines correspond to the roughness exponents for a transient MBE regime ($\alpha\approx 1.141(9)$) and EW class in $d=1$ ($\alpha= 0.5$) that are associated with the scaling for $q\ll  0$ and $q \gg 0$, respectively. The shaded region represents the values of $q$ consistent with MBE regime. The inset illustrates the corresponding singularity spectrum.}
	\label{fig:MFRSOS}
\end{figure*}

MF-ODFA is based on a generalization of the ODFA described in Ref.~\cite{Luis17}. Let us consider a discrete interface of height $h$. We divide the interface into $N_{r}$  windows of equal sizes $r$. For each window  (labeled by $1 \leq \nu \leq N_r$), a   polynomial of degree $n$, $P_{\nu}^{n}(x:A_{\nu}^{(0)}, A_{\nu}^{(1)}, ..., A_{\nu}^{(n)})$,  is obtained using the least-squares' method~\cite{lsq} providing the the coefficients $A_{x}^{(0)}, A_{x}^{(1)}, \cdots, A_{x}^{(n)}$. For each position $x$ within the window, the minimal distance between the interface height $h(x)$ and the polynomial regression, represented by $\delta_{x}^{n}$, is calculated, as schematically shown in the Fig. \ref{fig:models_prof}(b).

The $q$th-order dispersion of a window $\nu$ with size $r$ is calculated as follows:
\begin{equation}
\mathcal{F}_{q}^{n}(r,\nu) = \left\langle (\delta_{x} ^{n})^{2} \right\rangle_{\nu} ^{q/2},
\end{equation}
where $\langle \cdot \rangle_{\nu}$ is average over the window $\nu$. Note that $q$ serves as a parameter to investigate the complexity of the interfaces across different scales, enhancing the shorter ($q<0$) and larger ($q>0$) height fluctuations in an interface. The $q$th-order  fluctuations for the different sizes are calculated as
\begin{equation}
\label{eq:fl1}
\omega_{q}^{n}(r) = \left[ \left\langle \mathcal{F}_{q}^{n}(r,\nu)\right\rangle_r \right]^{1/q}.
\end{equation}
One expects the scaling $\omega_{q}^{n}(r)  \sim r^{\alpha_{l}(q)}$. There exist different methods for extracting $\alpha_{l}(q)$, for example, through the generalized height-height correlation function~\cite{Barabasi}, or the multifractal detrended fluctuation analysis (MF-DFA)~\cite{Jan02}. Indeed, Eq. (\ref{eq:fl1}) yields the local roughness of a undetrended interface when $q =2$ and $P_{x}^{0} = A_{x}^{0} = \overline{h}_r$. 

The multifractal exponent $\tau(q)$ is connected with the generalized local roughness exponent, $\alpha_l(q)$, through of the relation ~\cite{Kimiagar_2009} 
\begin{equation}
\label{tauq}
\tau(q) = q\alpha_{l}(q) -1.
\end{equation}

Another way to describe the multifractality is the singularity spectrum $f(\Theta)$, where $\Theta$ is the singularity strength~\cite{Jiang19}. These two variables are related through a Legendre transform~\cite{Halsey_86,Feder88,Jiang19} and described by the following relation
\begin{equation}
f (\Theta) = q\Theta - \tau (q)
\end{equation}
where
\begin{equation}
	\Theta = \frac{d \tau(q)}{d q}.
\end{equation}

The ODFA method is a consistent methodology for analyzing mounded interfaces~\cite{Luis17,Luis19}. Furthermore, MF-ODFA, which also works for self-affine models, confirms the mono-fractal feature of a well-known restricted solid-on-solid (RSOS) model ~\cite{Kim_89} that belongs to the Kadar-Parisi-Zhang class~\cite{Kardar86} (see Fig. S3 in the SM).

\textit{Results and discussion}. For a multifractal structure, a typical $\tau(q)$ approaches asymptotically $\tau(q)=\alpha_{l}^\mathrm{max} q$ when $q\rightarrow -\infty$ and  $\tau(q)=\alpha_{l}^\mathrm{min} q$ when $q\rightarrow+\infty$~\cite{Feder88}. These asymptotic regimes are smoothly connected by $\tau(q)$. In Fig.~\ref{fig:MFRSOS}(a), the $q$-th order fluctuations as a function of various scale lengths are shown for $-10<q<20$. The corresponding local slope analyses, given by $\alpha_\mathrm{l}^{\mathrm{eff}}(q) \equiv d  \ln \omega_{q}^{n}(r) /d \ln r $, are presented in Fig.~\ref{fig:MFRSOS}(b) for the WV model with $n=2$ and $t=10^8$. Figures~\ref{fig:MFRSOS}(a) and ~\ref{fig:MFRSOS}(b) show different scaling regimes for $q \ll  0$ and $q \gg 0$, characterized by the plateaus of $\alpha_{l}^{\mathrm{eff}}(q)$ for length scales that are smaller or larger than the correlation length of the interface. So, the MF-ODFA method detects different scale regimes considering length scales that are smal or larger than the correlation length of the interface. This feature was not observed previously, including the MF-DFA, highlighting the advantage and efficiency of the MF-ODFA. 

For each $q$-value, a least square fitting in the range of $r$ where the plateau of $\alpha_{l}^{\mathrm{eff}}(q)$ is constant defines $\alpha_l (q)$, which is then used to compute $\tau(q)$ in Eq. (\ref{tauq}). The multifractal exponent shows a crossover between two scaling regimes for negative and positive  $q$ while the absence of scaling is found for $q$ close to 0. Figure~\ref{fig:MFRSOS}(c) depicts $\tau(q)$ versus $q$, confirming the presence of two linear segments for $q \ll 0$ and $q \gg 0$, pointing a bifractal structure where two self-affine scaling regimes occur at two distinct scales.  This bifractal feature is further supported by the inset of Fig.~\ref{fig:MFRSOS}(c), which indicates that the singularity exponent $\Theta$ concentrates around two values.

\begin{figure}[hbt]
 	\center 	
    \includegraphics [width=0.65\linewidth]{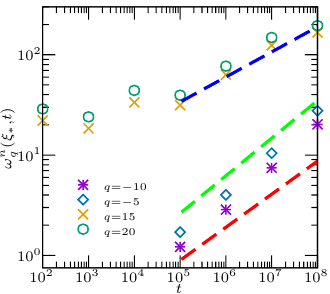}
 	\caption{(color online) Local roughness  $\omega_{q}^{n}(r=\xi_{*},t)$ vs. $t$, for $n=2$ and different $q$-values. The red, green, and blue dashed lines represent the slopes $\beta^{\mathrm{VLDS}} \approx 0.33$, $\beta^{\mathrm{MH}}=3/8$ and $\beta^{\mathrm{EW}}=0.5$. In the case of the VLDS class, the growth exponent is derived from a two-loop RG analysis \cite{Janssen97}. For $q=15$ [$q=20$], the effective growth exponent, calculated by least squares fitting of the numerical data, was $0.24(2)$ [$0.23(2)$]. For $q=-5$ [$q=-10$], the effective growth exponent was $0.404(8)$ [$0.40(1)$]. The correlation length $\xi_{*}(t)$ is defined by $\Gamma(\xi_{*},t)/\Gamma(0) = 0.3$.}
 	\label{fig:ExpRSOSn}
 \end{figure}

Our study regarding the exponent $\tau(q)$ demonstrates an effective local roughness exponent in agreement with the MBE growth across $q\ll 0$. On the one hand, a two-loop renormalization group theory for nonlinear MBE theory, given by Eq.~\eqref{eq:sto_eq} with $\nu=0$,  yields the scaling  exponents for the VLDS equation as $\alpha^{\mathrm{VLDS}}(d) = \left({4-d}\right)/{3}-\zeta(d)$ , and $z^{\mathrm{VLDS}}(d) = (d + 8)/3-2\zeta(d)$~\cite{Janssen97},
where  $\zeta(d)=0.01361\left(2-{d}/{2}\right)^2$. Hence, for $d = 1$, $\zeta = 0.0306$, yielding $\alpha^{\mathrm{VLDS}} \approx 0.97$ and $\beta^{\mathrm{VLDS}}= \alpha^{\mathrm{VLDS}}/z^{\mathrm{VLDS}} \approx 0.33$. Recent studies using the ODFA technique have shown that the local roughness exponents numerically calculated for a variety of models, where diffusive dynamics were dominant, exhibited values very similar to the global roughness exponent for the VLDS class in $d=1$ and $2$~\cite{Luis19, Luis17}. On the other hand, the linear MBE theory stated by the MH equation provides $\alpha=1.5$ and $\beta=0.375$. The scaling analysis for $q<0$ are consistent with $\alpha=1.141(9)$ which lays between the linear and nonlinear MBE theories. So, we conjecture that the short-scale regime is a crossover between linear and nonlinear MBE universality classes. A crossover for the leading VLDS class is expected for even more asymptotic regimes.

For $q > 0$, the scaling properties associated with long-wavelength fluctuations are in agreement with the EW regime, and consistent with previous observations~\cite{Xuzhou15,Kotrla96,xun12}. Interestingly, EW scaling for long-wavelength fluctuations was observed at a time ($t=10^8$), that is one order of magnitude smaller than the recent analysis of WV models~\cite{xun12}. However, as short wavelength fluctuations scale as in the transient between MH and VLDS classes (i.e., MBE growth), we present here the first evidence of a bifractal pattern for the WV model, which contrasts with the hypothesis of monofractality in this model.

We also investigate the time scaling in the growth regime of the local roughness, computing the roughness at length scales equivalent to a correlation length determined by the condition $\Gamma(\xi_{*},t)/\Gamma(0) = 0.3$; see Fig. S2 in the SM. The reliability of the definition of $\xi_{*}$ to calculate the dynamic exponent, $z$, was demonstrated also for several interface growth models~\cite{Reis2015,PRE2023}. The choice of this length scale is justified by the observation of concomitant local roughness exponents corresponding to MBE growth and EW class for $q<0$ and $q>0$, respectively, when $r = \xi_{*}$; see Fig.~\ref{fig:MFRSOS}(b). We calculate $\omega_{q}^{n}(r=\xi_{*},t)$, with $n=2$, as function of time for both negative and positive $q$-values as shown in Figure \ref{fig:ExpRSOSn}. For long times $t \gtrsim 10^{5}$, two temporal scaling regimes are observed, depending on the sign of $q$. For negative $q$, the effective growth exponent $\beta=0.40(1)$ is close to linear MBE theory  $\beta=0.375$, being a consistent scaling regime before the dominance of the nonlinear terms. Indeed, the growth exponents of linear and nonlinear MBE universality classes are relatively close such that resolving between them demands scaling above those we were not able to achieve. For positive $q$, the growth exponent $\beta=0.24(2)$ is fully consistent with the EW universality class.

\textit{Summary and Conclusions}. In summary, we introduced the MF-ODFA method, a powerful tool for studying multifractality in film growth models, resulting in interfaces with characteristic lengths over long time intervals. We applied our method to the WV model in $d=1$, demonstrating a bifractal morphology, contradicting the hypothesis that the model, at least within the reported times, is monofractal. For short-scale lengths, the MF-ODFA methods provide scaling exponents consistent with the MBE universality class whereas the EW exponents are observed for large scales, in agreement with the conjecture that this models belong to the latter class in the hydrodynamic limit~\cite{Kotrla96}. We emphisize that our method can be effectively applied to analyze other models exhibiting long transient mounded morphologies, where determining the universality class remains uncovered.

\textit{Acknowledgements}. This work was supported by the Conselho Nacional de Desenvolvimento Cient\'{i}fico e Tecnol\'{o}gico (CNPq), Grant Nos. 305688/2023-5 (TAdA) and 310984/2023-8 (SCF). EEML is grateful for funding from CNPq (01300.008811/2022-51) and Funda\c{c}\~{a}o de Amparo \`{a} Pesquisa do Estado do Rio de Janeiro (SEI-260003/005791/2022). The authors would like to thank Jos\'{e} G. V. Miranda and F\'{a}bio Aar\~{a}o Reis for fruitful discussions.

%\bibliography{Refs_MF_WV.bib}

\begin{thebibliography}{49}%
\makeatletter
\providecommand \@ifxundefined [1]{%
 \@ifx{#1\undefined}
}%
\providecommand \@ifnum [1]{%
 \ifnum #1\expandafter \@firstoftwo
 \else \expandafter \@secondoftwo
 \fi
}%
\providecommand \@ifx [1]{%
 \ifx #1\expandafter \@firstoftwo
 \else \expandafter \@secondoftwo
 \fi
}%
\providecommand \natexlab [1]{#1}%
\providecommand \enquote  [1]{``#1''}%
\providecommand \bibnamefont  [1]{#1}%
\providecommand \bibfnamefont [1]{#1}%
\providecommand \citenamefont [1]{#1}%
\providecommand \href@noop [0]{\@secondoftwo}%
\providecommand \href [0]{\begingroup \@sanitize@url \@href}%
\providecommand \@href[1]{\@@startlink{#1}\@@href}%
\providecommand \@@href[1]{\endgroup#1\@@endlink}%
\providecommand \@sanitize@url [0]{\catcode `\\12\catcode `\$12\catcode `\&12\catcode `\#12\catcode `\^12\catcode `\_12\catcode `\%12\relax}%
\providecommand \@@startlink[1]{}%
\providecommand \@@endlink[0]{}%
\providecommand \url  [0]{\begingroup\@sanitize@url \@url }%
\providecommand \@url [1]{\endgroup\@href {#1}{\urlprefix }}%
\providecommand \urlprefix  [0]{URL }%
\providecommand \Eprint [0]{\href }%
\providecommand \doibase [0]{https://doi.org/}%
\providecommand \selectlanguage [0]{\@gobble}%
\providecommand \bibinfo  [0]{\@secondoftwo}%
\providecommand \bibfield  [0]{\@secondoftwo}%
\providecommand \translation [1]{[#1]}%
\providecommand \BibitemOpen [0]{}%
\providecommand \bibitemStop [0]{}%
\providecommand \bibitemNoStop [0]{.\EOS\space}%
\providecommand \EOS [0]{\spacefactor3000\relax}%
\providecommand \BibitemShut  [1]{\csname bibitem#1\endcsname}%
\let\auto@bib@innerbib\@empty
%</preamble>
\bibitem [{\citenamefont {Krug}(1997)}]{Krug97}%
  \BibitemOpen
  \bibfield  {author} {\bibinfo {author} {\bibfnamefont {J.}~\bibnamefont {Krug}},\ }\bibfield  {title} {\bibinfo {title} {{O}rigins of scale invariance in growth processes},\ }\href {https://doi.org/10.1080/00018739700101498} {\bibfield  {journal} {\bibinfo  {journal} {Advances in Physics}\ }\textbf {\bibinfo {volume} {46}},\ \bibinfo {pages} {139} (\bibinfo {year} {1997})}\BibitemShut {NoStop}%
\bibitem [{\citenamefont {Ohring}(2002)}]{Ohring2002}%
  \BibitemOpen
  \bibfield  {author} {\bibinfo {author} {\bibfnamefont {M.}~\bibnamefont {Ohring}},\ }\href@noop {} {\emph {\bibinfo {title} {Materials Science of Thin Films: Depositon \& Structure}}}\ (\bibinfo  {publisher} {Elsevier Science},\ \bibinfo {year} {2002})\BibitemShut {NoStop}%
\bibitem [{\citenamefont {Evans}\ \emph {et~al.}(2006)\citenamefont {Evans}, \citenamefont {Thiel},\ and\ \citenamefont {Bartelt}}]{Evans2006}%
  \BibitemOpen
  \bibfield  {author} {\bibinfo {author} {\bibfnamefont {J.}~\bibnamefont {Evans}}, \bibinfo {author} {\bibfnamefont {P.}~\bibnamefont {Thiel}},\ and\ \bibinfo {author} {\bibfnamefont {M.}~\bibnamefont {Bartelt}},\ }\bibfield  {title} {\bibinfo {title} {Morphological evolution during epitaxial thin film growth: Formation of 2d islands and 3d mounds},\ }\href {https://doi.org/https://doi.org/10.1016/j.surfrep.2005.08.004} {\bibfield  {journal} {\bibinfo  {journal} {Surface Science Reports}\ }\textbf {\bibinfo {volume} {61}},\ \bibinfo {pages} {1} (\bibinfo {year} {2006})}\BibitemShut {NoStop}%
\bibitem [{\citenamefont {Quastel}\ and\ \citenamefont {Spohn}(2015)}]{Spohn2015}%
  \BibitemOpen
  \bibfield  {author} {\bibinfo {author} {\bibfnamefont {J.}~\bibnamefont {Quastel}}\ and\ \bibinfo {author} {\bibfnamefont {H.}~\bibnamefont {Spohn}},\ }\bibfield  {title} {\bibinfo {title} {The one-dimensional kpz equation and its universality class},\ }\href {https://doi.org/10.1007/s10955-015-1250-9} {\bibfield  {journal} {\bibinfo  {journal} {Journal of Statistical Physics}\ }\textbf {\bibinfo {volume} {160}},\ \bibinfo {pages} {965} (\bibinfo {year} {2015})}\BibitemShut {NoStop}%
\bibitem [{\citenamefont {Gomes}\ \emph {et~al.}(2019)\citenamefont {Gomes}, \citenamefont {Penna},\ and\ \citenamefont {Oliveira}}]{Oliveira2019}%
  \BibitemOpen
  \bibfield  {author} {\bibinfo {author} {\bibfnamefont {W.~P.}\ \bibnamefont {Gomes}}, \bibinfo {author} {\bibfnamefont {A.~L.~A.}\ \bibnamefont {Penna}},\ and\ \bibinfo {author} {\bibfnamefont {F.~A.}\ \bibnamefont {Oliveira}},\ }\bibfield  {title} {\bibinfo {title} {From cellular automata to growth dynamics: The kardar-parisi-zhang universality class},\ }\href {https://doi.org/10.1103/PhysRevE.100.020101} {\bibfield  {journal} {\bibinfo  {journal} {Phys. Rev. E}\ }\textbf {\bibinfo {volume} {100}},\ \bibinfo {pages} {020101} (\bibinfo {year} {2019})}\BibitemShut {NoStop}%
\bibitem [{\citenamefont {Mallio}\ and\ \citenamefont {{Aar\~{a}o Reis}}(2022)}]{Mallio2022}%
  \BibitemOpen
  \bibfield  {author} {\bibinfo {author} {\bibfnamefont {D.~O.}\ \bibnamefont {Mallio}}\ and\ \bibinfo {author} {\bibfnamefont {F.}~\bibnamefont {{Aar\~{a}o Reis}}},\ }\bibfield  {title} {\bibinfo {title} {Short length scale fluctuations in lattice growth models},\ }\href {https://doi.org/https://doi.org/10.1016/j.physa.2022.127178} {\bibfield  {journal} {\bibinfo  {journal} {Physica A: Statistical Mechanics and its Applications}\ }\textbf {\bibinfo {volume} {596}},\ \bibinfo {pages} {127178} (\bibinfo {year} {2022})}\BibitemShut {NoStop}%
\bibitem [{\citenamefont {Chame}\ and\ \citenamefont {{Aar\~{a}o Reis}}(2004)}]{Chame2004}%
  \BibitemOpen
  \bibfield  {author} {\bibinfo {author} {\bibfnamefont {A.}~\bibnamefont {Chame}}\ and\ \bibinfo {author} {\bibfnamefont {F.}~\bibnamefont {{Aar\~{a}o Reis}}},\ }\bibfield  {title} {\bibinfo {title} {Scaling of local interface width of statistical growth models},\ }\href {https://doi.org/https://doi.org/10.1016/j.susc.2004.01.048} {\bibfield  {journal} {\bibinfo  {journal} {Surface Science}\ }\textbf {\bibinfo {volume} {553}},\ \bibinfo {pages} {145} (\bibinfo {year} {2004})}\BibitemShut {NoStop}%
\bibitem [{\citenamefont {Mozo~Luis}\ \emph {et~al.}(2023)\citenamefont {Mozo~Luis}, \citenamefont {Oliveira},\ and\ \citenamefont {de~Assis}}]{PRE2023}%
  \BibitemOpen
  \bibfield  {author} {\bibinfo {author} {\bibfnamefont {E.~E.}\ \bibnamefont {Mozo~Luis}}, \bibinfo {author} {\bibfnamefont {F.~A.}\ \bibnamefont {Oliveira}},\ and\ \bibinfo {author} {\bibfnamefont {T.~A.}\ \bibnamefont {de~Assis}},\ }\bibfield  {title} {\bibinfo {title} {Accessibility of the surface fractal dimension during film growth},\ }\href {https://doi.org/10.1103/PhysRevE.107.034802} {\bibfield  {journal} {\bibinfo  {journal} {Phys. Rev. E}\ }\textbf {\bibinfo {volume} {107}},\ \bibinfo {pages} {034802} (\bibinfo {year} {2023})}\BibitemShut {NoStop}%
\bibitem [{\citenamefont {Wolf}\ and\ \citenamefont {Villain}(1990)}]{Wolf_90}%
  \BibitemOpen
  \bibfield  {author} {\bibinfo {author} {\bibfnamefont {D.~E.}\ \bibnamefont {Wolf}}\ and\ \bibinfo {author} {\bibfnamefont {J.}~\bibnamefont {Villain}},\ }\bibfield  {title} {\bibinfo {title} {Growth with surface diffusion},\ }\href {https://doi.org/10.1209/0295-5075/13/5/002} {\bibfield  {journal} {\bibinfo  {journal} {Europhysics Letters ({EPL})}\ }\textbf {\bibinfo {volume} {13}},\ \bibinfo {pages} {389} (\bibinfo {year} {1990})}\BibitemShut {NoStop}%
\bibitem [{\citenamefont {Schroeder}\ \emph {et~al.}(1993)\citenamefont {Schroeder}, \citenamefont {Siegert}, \citenamefont {Wolf}, \citenamefont {Shore},\ and\ \citenamefont {Plischke}}]{Schroeder93}%
  \BibitemOpen
  \bibfield  {author} {\bibinfo {author} {\bibfnamefont {M.}~\bibnamefont {Schroeder}}, \bibinfo {author} {\bibfnamefont {M.}~\bibnamefont {Siegert}}, \bibinfo {author} {\bibfnamefont {D.~E.}\ \bibnamefont {Wolf}}, \bibinfo {author} {\bibfnamefont {J.~D.}\ \bibnamefont {Shore}},\ and\ \bibinfo {author} {\bibfnamefont {M.}~\bibnamefont {Plischke}},\ }\bibfield  {title} {\bibinfo {title} {Scaling of growing surfaces with large local slopes},\ }\href {https://doi.org/10.1209/0295-5075/24/7/010} {\bibfield  {journal} {\bibinfo  {journal} {Europhysics Letters ({EPL})}\ }\textbf {\bibinfo {volume} {24}},\ \bibinfo {pages} {563} (\bibinfo {year} {1993})}\BibitemShut {NoStop}%
\bibitem [{\citenamefont {S\ifmmode~\check{}\else \v{}\fi{}milauer}\ and\ \citenamefont {Kotrla}(1994)}]{Pavel94}%
  \BibitemOpen
  \bibfield  {author} {\bibinfo {author} {\bibfnamefont {P.}~\bibnamefont {S\ifmmode~\check{}\else \v{}\fi{}milauer}}\ and\ \bibinfo {author} {\bibfnamefont {M.}~\bibnamefont {Kotrla}},\ }\bibfield  {title} {\bibinfo {title} {Crossover effects in the wolf-villain model of epitaxial growth in 1+1 and 2+1 dimensions},\ }\href {https://doi.org/10.1103/PhysRevB.49.5769} {\bibfield  {journal} {\bibinfo  {journal} {Phys. Rev. B}\ }\textbf {\bibinfo {volume} {49}},\ \bibinfo {pages} {5769} (\bibinfo {year} {1994})}\BibitemShut {NoStop}%
\bibitem [{\citenamefont {P\ifmmode~\check{r}\else \v{r}\fi{}edota}\ and\ \citenamefont {Kotrla}(1996)}]{Kotrla96}%
  \BibitemOpen
  \bibfield  {author} {\bibinfo {author} {\bibfnamefont {M.}~\bibnamefont {P\ifmmode~\check{r}\else \v{r}\fi{}edota}}\ and\ \bibinfo {author} {\bibfnamefont {M.}~\bibnamefont {Kotrla}},\ }\bibfield  {title} {\bibinfo {title} {Stochastic equations for simple discrete models of epitaxial growth},\ }\href {https://doi.org/10.1103/PhysRevE.54.3933} {\bibfield  {journal} {\bibinfo  {journal} {Phys. Rev. E}\ }\textbf {\bibinfo {volume} {54}},\ \bibinfo {pages} {3933} (\bibinfo {year} {1996})}\BibitemShut {NoStop}%
\bibitem [{\citenamefont {Huang}\ and\ \citenamefont {Gu}(1996)}]{Huang96}%
  \BibitemOpen
  \bibfield  {author} {\bibinfo {author} {\bibfnamefont {Z.-F.}\ \bibnamefont {Huang}}\ and\ \bibinfo {author} {\bibfnamefont {B.-L.}\ \bibnamefont {Gu}},\ }\bibfield  {title} {\bibinfo {title} {Growth equations for the wolf-villain and das sarma-tamborenea models of molecular-beam epitaxy},\ }\href {https://doi.org/10.1103/PhysRevE.54.5935} {\bibfield  {journal} {\bibinfo  {journal} {Phys. Rev. E}\ }\textbf {\bibinfo {volume} {54}},\ \bibinfo {pages} {5935} (\bibinfo {year} {1996})}\BibitemShut {NoStop}%
\bibitem [{\citenamefont {Punyindu}\ and\ \citenamefont {Das~Sarma}(1998)}]{Punyindu98}%
  \BibitemOpen
  \bibfield  {author} {\bibinfo {author} {\bibfnamefont {P.}~\bibnamefont {Punyindu}}\ and\ \bibinfo {author} {\bibfnamefont {S.}~\bibnamefont {Das~Sarma}},\ }\bibfield  {title} {\bibinfo {title} {Noise reduction and universality in limited-mobility models of nonequilibrium growth},\ }\href {https://doi.org/10.1103/PhysRevE.57.R4863} {\bibfield  {journal} {\bibinfo  {journal} {Phys. Rev. E}\ }\textbf {\bibinfo {volume} {57}},\ \bibinfo {pages} {R4863} (\bibinfo {year} {1998})}\BibitemShut {NoStop}%
\bibitem [{\citenamefont {Das~Sarma}\ \emph {et~al.}(2002)\citenamefont {Das~Sarma}, \citenamefont {Chatraphorn},\ and\ \citenamefont {Toroczkai}}]{Punyindu02}%
  \BibitemOpen
  \bibfield  {author} {\bibinfo {author} {\bibfnamefont {S.}~\bibnamefont {Das~Sarma}}, \bibinfo {author} {\bibfnamefont {P.~P.}\ \bibnamefont {Chatraphorn}},\ and\ \bibinfo {author} {\bibfnamefont {Z.}~\bibnamefont {Toroczkai}},\ }\bibfield  {title} {\bibinfo {title} {Universality class of discrete solid-on-solid limited mobility nonequilibrium growth models for kinetic surface roughening},\ }\href {https://doi.org/10.1103/PhysRevE.65.036144} {\bibfield  {journal} {\bibinfo  {journal} {Phys. Rev. E}\ }\textbf {\bibinfo {volume} {65}},\ \bibinfo {pages} {036144} (\bibinfo {year} {2002})}\BibitemShut {NoStop}%
\bibitem [{\citenamefont {Haselwandter}\ and\ \citenamefont {Vvedensky}(2007)}]{Haselwandter07}%
  \BibitemOpen
  \bibfield  {author} {\bibinfo {author} {\bibfnamefont {C.~A.}\ \bibnamefont {Haselwandter}}\ and\ \bibinfo {author} {\bibfnamefont {D.~D.}\ \bibnamefont {Vvedensky}},\ }\bibfield  {title} {\bibinfo {title} {Multiscale theory of fluctuating interfaces: Renormalization of atomistic models},\ }\href {https://doi.org/10.1103/PhysRevLett.98.046102} {\bibfield  {journal} {\bibinfo  {journal} {Phys. Rev. Lett.}\ }\textbf {\bibinfo {volume} {98}},\ \bibinfo {pages} {046102} (\bibinfo {year} {2007})}\BibitemShut {NoStop}%
\bibitem [{\citenamefont {Haselwandter}\ and\ \citenamefont {Vvedensky}(2008)}]{Haselwandter08}%
  \BibitemOpen
  \bibfield  {author} {\bibinfo {author} {\bibfnamefont {C.~A.}\ \bibnamefont {Haselwandter}}\ and\ \bibinfo {author} {\bibfnamefont {D.~D.}\ \bibnamefont {Vvedensky}},\ }\bibfield  {title} {\bibinfo {title} {Renormalization of stochastic lattice models: Epitaxial surfaces},\ }\href {https://doi.org/10.1103/PhysRevE.77.061129} {\bibfield  {journal} {\bibinfo  {journal} {Phys. Rev. E}\ }\textbf {\bibinfo {volume} {77}},\ \bibinfo {pages} {061129} (\bibinfo {year} {2008})}\BibitemShut {NoStop}%
\bibitem [{\citenamefont {Pereira}\ \emph {et~al.}(2019)\citenamefont {Pereira}, \citenamefont {Alves},\ and\ \citenamefont {Ferreira}}]{Ferreira2019}%
  \BibitemOpen
  \bibfield  {author} {\bibinfo {author} {\bibfnamefont {A.~J.}\ \bibnamefont {Pereira}}, \bibinfo {author} {\bibfnamefont {S.~G.}\ \bibnamefont {Alves}},\ and\ \bibinfo {author} {\bibfnamefont {S.~C.}\ \bibnamefont {Ferreira}},\ }\bibfield  {title} {\bibinfo {title} {Effects of a kinetic barrier on limited-mobility interface growth models},\ }\href {https://doi.org/10.1103/PhysRevE.99.042802} {\bibfield  {journal} {\bibinfo  {journal} {Phys. Rev. E}\ }\textbf {\bibinfo {volume} {99}},\ \bibinfo {pages} {042802} (\bibinfo {year} {2019})}\BibitemShut {NoStop}%
\bibitem [{\citenamefont {Xun}\ \emph {et~al.}(2012)\citenamefont {Xun}, \citenamefont {Tang}, \citenamefont {Han}, \citenamefont {Xia}, \citenamefont {Hao},\ and\ \citenamefont {Li}}]{xun12}%
  \BibitemOpen
  \bibfield  {author} {\bibinfo {author} {\bibfnamefont {Z.}~\bibnamefont {Xun}}, \bibinfo {author} {\bibfnamefont {G.}~\bibnamefont {Tang}}, \bibinfo {author} {\bibfnamefont {K.}~\bibnamefont {Han}}, \bibinfo {author} {\bibfnamefont {H.}~\bibnamefont {Xia}}, \bibinfo {author} {\bibfnamefont {D.}~\bibnamefont {Hao}},\ and\ \bibinfo {author} {\bibfnamefont {Y.}~\bibnamefont {Li}},\ }\bibfield  {title} {\bibinfo {title} {Asymptotic dynamic scaling behavior of the (1+1)-dimensional wolf-villain model},\ }\href {https://doi.org/10.1103/PhysRevE.85.041126} {\bibfield  {journal} {\bibinfo  {journal} {Phys. Rev. E}\ }\textbf {\bibinfo {volume} {85}},\ \bibinfo {pages} {041126} (\bibinfo {year} {2012})}\BibitemShut {NoStop}%
\bibitem [{\citenamefont {Family}\ and\ \citenamefont {Vicsek}(1985)}]{Family85}%
  \BibitemOpen
  \bibfield  {author} {\bibinfo {author} {\bibfnamefont {F.}~\bibnamefont {Family}}\ and\ \bibinfo {author} {\bibfnamefont {T.}~\bibnamefont {Vicsek}},\ }\bibfield  {title} {\bibinfo {title} {Scaling of the active zone in the eden process on percolation networks and the ballistic deposition model},\ }\href {https://doi.org/10.1088/0305-4470/18/2/005} {\bibfield  {journal} {\bibinfo  {journal} {Journal of Physics A: Mathematical and General}\ }\textbf {\bibinfo {volume} {18}},\ \bibinfo {pages} {L75} (\bibinfo {year} {1985})}\BibitemShut {NoStop}%
\bibitem [{\citenamefont {Barab{\'a}si}\ and\ \citenamefont {Stanley}(1995)}]{Barabasi}%
  \BibitemOpen
  \bibfield  {author} {\bibinfo {author} {\bibfnamefont {A.~L.}\ \bibnamefont {Barab{\'a}si}}\ and\ \bibinfo {author} {\bibfnamefont {H.~E.}\ \bibnamefont {Stanley}},\ }\href@noop {} {\emph {\bibinfo {title} {{F}ractal concepts in surface growth}}}\ (\bibinfo  {publisher} {Cambridge university press},\ \bibinfo {year} {1995})\BibitemShut {NoStop}%
\bibitem [{\citenamefont {Aar\~ao Reis}(2013)}]{Reis13}%
  \BibitemOpen
  \bibfield  {author} {\bibinfo {author} {\bibfnamefont {F.~D.~A.}\ \bibnamefont {Aar\~ao Reis}},\ }\bibfield  {title} {\bibinfo {title} {Normal dynamic scaling in the class of the nonlinear molecular-beam-epitaxy equation},\ }\href {https://doi.org/10.1103/PhysRevE.88.022128} {\bibfield  {journal} {\bibinfo  {journal} {Phys. Rev. E}\ }\textbf {\bibinfo {volume} {88}},\ \bibinfo {pages} {022128} (\bibinfo {year} {2013})}\BibitemShut {NoStop}%
\bibitem [{\citenamefont {L\'opez}\ \emph {et~al.}(1997)\citenamefont {L\'opez}, \citenamefont {Rodr\'iguez},\ and\ \citenamefont {Cuerno}}]{Juan97}%
  \BibitemOpen
  \bibfield  {author} {\bibinfo {author} {\bibfnamefont {J.~M.}\ \bibnamefont {L\'opez}}, \bibinfo {author} {\bibfnamefont {M.~A.}\ \bibnamefont {Rodr\'iguez}},\ and\ \bibinfo {author} {\bibfnamefont {R.}~\bibnamefont {Cuerno}},\ }\bibfield  {title} {\bibinfo {title} {Power spectrum scaling in anomalous kinetic roughening of surfaces},\ }\href {https://doi.org/https://doi.org/10.1016/S0378-4371(97)00375-0} {\bibfield  {journal} {\bibinfo  {journal} {Physica A: Statistical Mechanics and its Applications}\ }\textbf {\bibinfo {volume} {246}},\ \bibinfo {pages} {329} (\bibinfo {year} {1997})}\BibitemShut {NoStop}%
\bibitem [{\citenamefont {L\'opez}(1999)}]{Juan99}%
  \BibitemOpen
  \bibfield  {author} {\bibinfo {author} {\bibfnamefont {J.~M.}\ \bibnamefont {L\'opez}},\ }\bibfield  {title} {\bibinfo {title} {Scaling approach to calculate critical exponents in anomalous surface roughening},\ }\href {https://doi.org/10.1103/PhysRevLett.83.4594} {\bibfield  {journal} {\bibinfo  {journal} {Phys. Rev. Lett.}\ }\textbf {\bibinfo {volume} {83}},\ \bibinfo {pages} {4594} (\bibinfo {year} {1999})}\BibitemShut {NoStop}%
\bibitem [{\citenamefont {L\'opez}\ \emph {et~al.}(2005)\citenamefont {L\'opez}, \citenamefont {Castro},\ and\ \citenamefont {Gallego}}]{Juan05}%
  \BibitemOpen
  \bibfield  {author} {\bibinfo {author} {\bibfnamefont {J.~M.}\ \bibnamefont {L\'opez}}, \bibinfo {author} {\bibfnamefont {M.}~\bibnamefont {Castro}},\ and\ \bibinfo {author} {\bibfnamefont {R.}~\bibnamefont {Gallego}},\ }\bibfield  {title} {\bibinfo {title} {Scaling of local slopes, conservation laws, and anomalous roughening in surface growth},\ }\href {https://doi.org/10.1103/PhysRevLett.94.166103} {\bibfield  {journal} {\bibinfo  {journal} {Phys. Rev. Lett.}\ }\textbf {\bibinfo {volume} {94}},\ \bibinfo {pages} {166103} (\bibinfo {year} {2005})}\BibitemShut {NoStop}%
\bibitem [{\citenamefont {Edwards}\ and\ \citenamefont {Wilkinson}(1982)}]{Edwards82}%
  \BibitemOpen
  \bibfield  {author} {\bibinfo {author} {\bibfnamefont {S.~F.}\ \bibnamefont {Edwards}}\ and\ \bibinfo {author} {\bibfnamefont {D.~R.}\ \bibnamefont {Wilkinson}},\ }\bibfield  {title} {\bibinfo {title} {The surface statistics of a granular aggregate},\ }\href {https://doi.org/10.1098/rspa.1982.0056} {\bibfield  {journal} {\bibinfo  {journal} {Proceedings of the Royal Society of London. A. Mathematical and Physical Sciences}\ }\textbf {\bibinfo {volume} {381}},\ \bibinfo {pages} {17} (\bibinfo {year} {1982})}\BibitemShut {NoStop}%
\bibitem [{\citenamefont {Mullins}(2004)}]{MH2004}%
  \BibitemOpen
  \bibfield  {author} {\bibinfo {author} {\bibfnamefont {W.~W.}\ \bibnamefont {Mullins}},\ }\bibfield  {title} {\bibinfo {title} {{Theory of Thermal Grooving}},\ }\href {https://doi.org/10.1063/1.1722742} {\bibfield  {journal} {\bibinfo  {journal} {Journal of Applied Physics}\ }\textbf {\bibinfo {volume} {28}},\ \bibinfo {pages} {333} (\bibinfo {year} {2004})}\BibitemShut {NoStop}%
\bibitem [{\citenamefont {Villain.}(1991)}]{Villain91}%
  \BibitemOpen
  \bibfield  {author} {\bibinfo {author} {\bibfnamefont {J.}~\bibnamefont {Villain.}},\ }\bibfield  {title} {\bibinfo {title} {Continuum models of crystal growth from atomic beams with and without desorption},\ }\href {https://doi.org/10.1051/jp1:1991114} {\bibfield  {journal} {\bibinfo  {journal} {J. Phys. I France.}\ }\textbf {\bibinfo {volume} {1}},\ \bibinfo {pages} {19} (\bibinfo {year} {1991})}\BibitemShut {NoStop}%
\bibitem [{\citenamefont {Lai}\ and\ \citenamefont {Das~Sarma}(1991)}]{Lai91}%
  \BibitemOpen
  \bibfield  {author} {\bibinfo {author} {\bibfnamefont {Z.-W.}\ \bibnamefont {Lai}}\ and\ \bibinfo {author} {\bibfnamefont {S.}~\bibnamefont {Das~Sarma}},\ }\bibfield  {title} {\bibinfo {title} {Kinetic growth with surface relaxation: Continuum versus atomistic models},\ }\href {https://doi.org/10.1103/PhysRevLett.66.2348} {\bibfield  {journal} {\bibinfo  {journal} {Phys. Rev. Lett.}\ }\textbf {\bibinfo {volume} {66}},\ \bibinfo {pages} {2348} (\bibinfo {year} {1991})}\BibitemShut {NoStop}%
\bibitem [{\citenamefont {Costa}\ \emph {et~al.}(2003)\citenamefont {Costa}, \citenamefont {Euz\'{e}bio},\ and\ \citenamefont {{Aar\~{a}o Reis}}}]{Costa2003}%
  \BibitemOpen
  \bibfield  {author} {\bibinfo {author} {\bibfnamefont {B.}~\bibnamefont {Costa}}, \bibinfo {author} {\bibfnamefont {J.}~\bibnamefont {Euz\'{e}bio}},\ and\ \bibinfo {author} {\bibfnamefont {F.}~\bibnamefont {{Aar\~{a}o Reis}}},\ }\bibfield  {title} {\bibinfo {title} {Finite-size effects on the growth models of das sarma and tamborenea and wolf and villain},\ }\href {https://doi.org/https://doi.org/10.1016/S0378-4371(03)00581-8} {\bibfield  {journal} {\bibinfo  {journal} {Physica A: Statistical Mechanics and its Applications}\ }\textbf {\bibinfo {volume} {328}},\ \bibinfo {pages} {193} (\bibinfo {year} {2003})}\BibitemShut {NoStop}%
\bibitem [{\citenamefont {Xun}\ \emph {et~al.}(2010)\citenamefont {Xun}, \citenamefont {Tang}, \citenamefont {Han}, \citenamefont {Xia}, \citenamefont {Hao}, \citenamefont {Yang},\ and\ \citenamefont {Zhou}}]{Xun2010}%
  \BibitemOpen
  \bibfield  {author} {\bibinfo {author} {\bibfnamefont {Z.}~\bibnamefont {Xun}}, \bibinfo {author} {\bibfnamefont {G.}~\bibnamefont {Tang}}, \bibinfo {author} {\bibfnamefont {K.}~\bibnamefont {Han}}, \bibinfo {author} {\bibfnamefont {H.}~\bibnamefont {Xia}}, \bibinfo {author} {\bibfnamefont {D.}~\bibnamefont {Hao}}, \bibinfo {author} {\bibfnamefont {X.}~\bibnamefont {Yang}},\ and\ \bibinfo {author} {\bibfnamefont {W.}~\bibnamefont {Zhou}},\ }\bibfield  {title} {\bibinfo {title} {Extensive numerical study of the anomalous dynamic scaling of the wolf–villain model},\ }\href {https://doi.org/https://doi.org/10.1016/j.physa.2010.01.051} {\bibfield  {journal} {\bibinfo  {journal} {Physica A: Statistical Mechanics and its Applications}\ }\textbf {\bibinfo {volume} {389}},\ \bibinfo {pages} {2189} (\bibinfo {year} {2010})}\BibitemShut {NoStop}%
\bibitem [{\citenamefont {Zhao}\ \emph {et~al.}(2001)\citenamefont {Zhao}, \citenamefont {Wang},\ and\ \citenamefont {Lu}}]{Zhao2001}%
  \BibitemOpen
  \bibfield  {author} {\bibinfo {author} {\bibfnamefont {Y.}~\bibnamefont {Zhao}}, \bibinfo {author} {\bibfnamefont {G.-C.}\ \bibnamefont {Wang}},\ and\ \bibinfo {author} {\bibfnamefont {T.-M.}\ \bibnamefont {Lu}},\ }\bibfield  {title} {\bibinfo {title} {Characterization of amorphous and crystalline rough surface: Principles and applications},\ }\href@noop {} {\bibfield  {journal} {\bibinfo  {journal} {Characterization of Amorphous and Crystalline Rough Surface: Principles and Applications. Series: Experimental Methods in the Physical Sciences, ISBN: 9780124759848. Elsevier, vol. 37, pp. xv-xvii}\ }\textbf {\bibinfo {volume} {37}},\ \bibinfo {pages} {15} (\bibinfo {year} {2001})}\BibitemShut {NoStop}%
\bibitem [{\citenamefont {Luis}\ \emph {et~al.}(2022)\citenamefont {Luis}, \citenamefont {de~Assis},\ and\ \citenamefont {Oliveira}}]{Mozo2022}%
  \BibitemOpen
  \bibfield  {author} {\bibinfo {author} {\bibfnamefont {E.~E.~M.}\ \bibnamefont {Luis}}, \bibinfo {author} {\bibfnamefont {T.~A.}\ \bibnamefont {de~Assis}},\ and\ \bibinfo {author} {\bibfnamefont {F.~A.}\ \bibnamefont {Oliveira}},\ }\bibfield  {title} {\bibinfo {title} {Unveiling the connection between the global roughness exponent and interface fractal dimension in {EW} and {KPZ} lattice models},\ }\href {https://doi.org/10.1088/1742-5468/ac7e3f} {\bibfield  {journal} {\bibinfo  {journal} {Journal of Statistical Mechanics: Theory and Experiment}\ }\textbf {\bibinfo {volume} {2022}},\ \bibinfo {pages} {083202} (\bibinfo {year} {2022})}\BibitemShut {NoStop}%
\bibitem [{\citenamefont {Luis}\ \emph {et~al.}(2017)\citenamefont {Luis}, \citenamefont {de~Assis},\ and\ \citenamefont {Ferreira}}]{Luis17}%
  \BibitemOpen
  \bibfield  {author} {\bibinfo {author} {\bibfnamefont {E.~E.~M.}\ \bibnamefont {Luis}}, \bibinfo {author} {\bibfnamefont {T.~A.}\ \bibnamefont {de~Assis}},\ and\ \bibinfo {author} {\bibfnamefont {S.~C.}\ \bibnamefont {Ferreira}},\ }\bibfield  {title} {\bibinfo {title} {Optimal detrended fluctuation analysis as a tool for the determination of the roughness exponent of the mounded surfaces},\ }\href {https://doi.org/10.1103/PhysRevE.95.042801} {\bibfield  {journal} {\bibinfo  {journal} {Phys. Rev. E}\ }\textbf {\bibinfo {volume} {95}},\ \bibinfo {pages} {042801} (\bibinfo {year} {2017})}\BibitemShut {NoStop}%
\bibitem [{\citenamefont {Luis}\ \emph {et~al.}(2019)\citenamefont {Luis}, \citenamefont {de~Assis}, \citenamefont {Ferreira},\ and\ \citenamefont {Andrade}}]{Luis19}%
  \BibitemOpen
  \bibfield  {author} {\bibinfo {author} {\bibfnamefont {E.~E.~M.}\ \bibnamefont {Luis}}, \bibinfo {author} {\bibfnamefont {T.~A.}\ \bibnamefont {de~Assis}}, \bibinfo {author} {\bibfnamefont {S.~C.}\ \bibnamefont {Ferreira}},\ and\ \bibinfo {author} {\bibfnamefont {R.~F.~S.}\ \bibnamefont {Andrade}},\ }\bibfield  {title} {\bibinfo {title} {Local roughness exponent in the nonlinear molecular-beam-epitaxy universality class in one dimension},\ }\href {https://doi.org/10.1103/PhysRevE.99.022801} {\bibfield  {journal} {\bibinfo  {journal} {Phys. Rev. E}\ }\textbf {\bibinfo {volume} {99}},\ \bibinfo {pages} {022801} (\bibinfo {year} {2019})}\BibitemShut {NoStop}%
\bibitem [{\citenamefont {{Ramana Murty}}\ and\ \citenamefont {Cooper}(2003)}]{Murty2003}%
  \BibitemOpen
  \bibfield  {author} {\bibinfo {author} {\bibfnamefont {M.}~\bibnamefont {{Ramana Murty}}}\ and\ \bibinfo {author} {\bibfnamefont {B.}~\bibnamefont {Cooper}},\ }\bibfield  {title} {\bibinfo {title} {Influence of step edge diffusion on surface morphology during epitaxy},\ }\href {https://doi.org/https://doi.org/10.1016/S0039-6028(03)00749-0} {\bibfield  {journal} {\bibinfo  {journal} {Surface Science}\ }\textbf {\bibinfo {volume} {539}},\ \bibinfo {pages} {91} (\bibinfo {year} {2003})}\BibitemShut {NoStop}%
\bibitem [{\citenamefont {Siniscalco}\ \emph {et~al.}(2013)\citenamefont {Siniscalco}, \citenamefont {Edely}, \citenamefont {Bardeau},\ and\ \citenamefont {Delorme}}]{mound}%
  \BibitemOpen
  \bibfield  {author} {\bibinfo {author} {\bibfnamefont {D.}~\bibnamefont {Siniscalco}}, \bibinfo {author} {\bibfnamefont {M.}~\bibnamefont {Edely}}, \bibinfo {author} {\bibfnamefont {J.-F.}\ \bibnamefont {Bardeau}},\ and\ \bibinfo {author} {\bibfnamefont {N.}~\bibnamefont {Delorme}},\ }\bibfield  {title} {\bibinfo {title} {Statistical analysis of mounded surfaces: Application to the evolution of ultrathin gold film morphology with deposition temperature},\ }\href {https://doi.org/10.1021/la304621k} {\bibfield  {journal} {\bibinfo  {journal} {Langmuir}\ }\textbf {\bibinfo {volume} {29}},\ \bibinfo {pages} {717} (\bibinfo {year} {2013})}\BibitemShut {NoStop}%
\bibitem [{\citenamefont {Leal}\ \emph {et~al.}(2011)\citenamefont {Leal}, \citenamefont {Ferreira},\ and\ \citenamefont {Ferreira}}]{Leal2011}%
  \BibitemOpen
  \bibfield  {author} {\bibinfo {author} {\bibfnamefont {F.~F.}\ \bibnamefont {Leal}}, \bibinfo {author} {\bibfnamefont {S.~C.}\ \bibnamefont {Ferreira}},\ and\ \bibinfo {author} {\bibfnamefont {S.~O.}\ \bibnamefont {Ferreira}},\ }\bibfield  {title} {\bibinfo {title} {Modelling of epitaxial film growth with an ehrlich–schwoebel barrier dependent on the step height},\ }\href {https://doi.org/10.1088/0953-8984/23/29/292201} {\bibfield  {journal} {\bibinfo  {journal} {Journal of Physics: Condensed Matter}\ }\textbf {\bibinfo {volume} {23}},\ \bibinfo {pages} {292201} (\bibinfo {year} {2011})}\BibitemShut {NoStop}%
\bibitem [{\citenamefont {Press}\ \emph {et~al.}(2007)\citenamefont {Press}, \citenamefont {Teukolsky}, \citenamefont {Vetterling},\ and\ \citenamefont {Flannery}}]{lsq}%
  \BibitemOpen
  \bibfield  {author} {\bibinfo {author} {\bibfnamefont {W.~H.}\ \bibnamefont {Press}}, \bibinfo {author} {\bibfnamefont {S.~A.}\ \bibnamefont {Teukolsky}}, \bibinfo {author} {\bibfnamefont {W.~T.}\ \bibnamefont {Vetterling}},\ and\ \bibinfo {author} {\bibfnamefont {B.~P.}\ \bibnamefont {Flannery}},\ }\href {https://books.google.com.br/books?id=1aAOdzK3FegC} {\emph {\bibinfo {title} {Numerical Recipes 3rd Edition: The Art of Scientific Computing}}},\ \bibinfo {edition} {3rd}\ ed.\ (\bibinfo  {publisher} {Cambridge University Press},\ \bibinfo {address} {New York, NY, USA},\ \bibinfo {year} {2007})\BibitemShut {NoStop}%
\bibitem [{\citenamefont {Kantelhardt}\ \emph {et~al.}(2002)\citenamefont {Kantelhardt}, \citenamefont {Zschiegner}, \citenamefont {Koscielny-Bunde}, \citenamefont {Havlin}, \citenamefont {Bunde},\ and\ \citenamefont {Stanley}}]{Jan02}%
  \BibitemOpen
  \bibfield  {author} {\bibinfo {author} {\bibfnamefont {J.~W.}\ \bibnamefont {Kantelhardt}}, \bibinfo {author} {\bibfnamefont {S.~A.}\ \bibnamefont {Zschiegner}}, \bibinfo {author} {\bibfnamefont {E.}~\bibnamefont {Koscielny-Bunde}}, \bibinfo {author} {\bibfnamefont {S.}~\bibnamefont {Havlin}}, \bibinfo {author} {\bibfnamefont {A.}~\bibnamefont {Bunde}},\ and\ \bibinfo {author} {\bibfnamefont {H.}~\bibnamefont {Stanley}},\ }\bibfield  {title} {\bibinfo {title} {Multifractal detrended fluctuation analysis of nonstationary time series},\ }\href {https://doi.org/https://doi.org/10.1016/S0378-4371(02)01383-3} {\bibfield  {journal} {\bibinfo  {journal} {Physica A: Statistical Mechanics and its Applications}\ }\textbf {\bibinfo {volume} {316}},\ \bibinfo {pages} {87} (\bibinfo {year} {2002})}\BibitemShut {NoStop}%
\bibitem [{\citenamefont {Kimiagar}\ \emph {et~al.}(2009)\citenamefont {Kimiagar}, \citenamefont {Movahed}, \citenamefont {Khorram}, \citenamefont {Sobhanian},\ and\ \citenamefont {Tabar}}]{Kimiagar_2009}%
  \BibitemOpen
  \bibfield  {author} {\bibinfo {author} {\bibfnamefont {S.}~\bibnamefont {Kimiagar}}, \bibinfo {author} {\bibfnamefont {M.~S.}\ \bibnamefont {Movahed}}, \bibinfo {author} {\bibfnamefont {S.}~\bibnamefont {Khorram}}, \bibinfo {author} {\bibfnamefont {S.}~\bibnamefont {Sobhanian}},\ and\ \bibinfo {author} {\bibfnamefont {M.~R.~R.}\ \bibnamefont {Tabar}},\ }\bibfield  {title} {\bibinfo {title} {Fractal analysis of discharge current fluctuations},\ }\href {https://doi.org/10.1088/1742-5468/2009/03/p03020} {\bibfield  {journal} {\bibinfo  {journal} {Journal of Statistical Mechanics: Theory and Experiment}\ }\textbf {\bibinfo {volume} {2009}},\ \bibinfo {pages} {P03020} (\bibinfo {year} {2009})}\BibitemShut {NoStop}%
\bibitem [{\citenamefont {Jiang}\ \emph {et~al.}(2019)\citenamefont {Jiang}, \citenamefont {Xie}, \citenamefont {Zhou},\ and\ \citenamefont {Sornette}}]{Jiang19}%
  \BibitemOpen
  \bibfield  {author} {\bibinfo {author} {\bibfnamefont {Z.-Q.}\ \bibnamefont {Jiang}}, \bibinfo {author} {\bibfnamefont {W.-J.}\ \bibnamefont {Xie}}, \bibinfo {author} {\bibfnamefont {W.-X.}\ \bibnamefont {Zhou}},\ and\ \bibinfo {author} {\bibfnamefont {D.}~\bibnamefont {Sornette}},\ }\bibfield  {title} {\bibinfo {title} {Multifractal analysis of financial markets: a review},\ }\href {https://doi.org/10.1088/1361-6633/ab42fb} {\bibfield  {journal} {\bibinfo  {journal} {Reports on Progress in Physics}\ }\textbf {\bibinfo {volume} {82}},\ \bibinfo {pages} {125901} (\bibinfo {year} {2019})}\BibitemShut {NoStop}%
\bibitem [{\citenamefont {Halsey}\ \emph {et~al.}(1986)\citenamefont {Halsey}, \citenamefont {Jensen}, \citenamefont {Kadanoff}, \citenamefont {Procaccia},\ and\ \citenamefont {Shraiman}}]{Halsey_86}%
  \BibitemOpen
  \bibfield  {author} {\bibinfo {author} {\bibfnamefont {T.~C.}\ \bibnamefont {Halsey}}, \bibinfo {author} {\bibfnamefont {M.~H.}\ \bibnamefont {Jensen}}, \bibinfo {author} {\bibfnamefont {L.~P.}\ \bibnamefont {Kadanoff}}, \bibinfo {author} {\bibfnamefont {I.}~\bibnamefont {Procaccia}},\ and\ \bibinfo {author} {\bibfnamefont {B.~I.}\ \bibnamefont {Shraiman}},\ }\bibfield  {title} {\bibinfo {title} {Fractal measures and their singularities: The characterization of strange sets},\ }\href {https://doi.org/10.1103/PhysRevA.33.1141} {\bibfield  {journal} {\bibinfo  {journal} {Phys. Rev. A}\ }\textbf {\bibinfo {volume} {33}},\ \bibinfo {pages} {1141} (\bibinfo {year} {1986})}\BibitemShut {NoStop}%
\bibitem [{\citenamefont {Feder.}(1988)}]{Feder88}%
  \BibitemOpen
  \bibfield  {author} {\bibinfo {author} {\bibfnamefont {J.}~\bibnamefont {Feder.}},\ }\href@noop {} {\emph {\bibinfo {title} {Fractals}}}\ (\bibinfo  {publisher} {Plenum Press, New York},\ \bibinfo {year} {1988})\BibitemShut {NoStop}%
\bibitem [{\citenamefont {Kim}\ and\ \citenamefont {Kosterlitz}(1989)}]{Kim_89}%
  \BibitemOpen
  \bibfield  {author} {\bibinfo {author} {\bibfnamefont {J.~M.}\ \bibnamefont {Kim}}\ and\ \bibinfo {author} {\bibfnamefont {J.~M.}\ \bibnamefont {Kosterlitz}},\ }\bibfield  {title} {\bibinfo {title} {Growth in a restricted solid-on-solid model},\ }\href {https://doi.org/10.1103/PhysRevLett.62.2289} {\bibfield  {journal} {\bibinfo  {journal} {Phys. Rev. Lett.}\ }\textbf {\bibinfo {volume} {62}},\ \bibinfo {pages} {2289} (\bibinfo {year} {1989})}\BibitemShut {NoStop}%
\bibitem [{\citenamefont {Kardar}\ \emph {et~al.}(1986)\citenamefont {Kardar}, \citenamefont {Parisi},\ and\ \citenamefont {Zhang}}]{Kardar86}%
  \BibitemOpen
  \bibfield  {author} {\bibinfo {author} {\bibfnamefont {M.}~\bibnamefont {Kardar}}, \bibinfo {author} {\bibfnamefont {G.}~\bibnamefont {Parisi}},\ and\ \bibinfo {author} {\bibfnamefont {Y.-C.}\ \bibnamefont {Zhang}},\ }\bibfield  {title} {\bibinfo {title} {Dynamic scaling of growing interfaces},\ }\href {https://doi.org/10.1103/PhysRevLett.56.889} {\bibfield  {journal} {\bibinfo  {journal} {Phys. Rev. Lett.}\ }\textbf {\bibinfo {volume} {56}},\ \bibinfo {pages} {889} (\bibinfo {year} {1986})}\BibitemShut {NoStop}%
\bibitem [{\citenamefont {Janssen}(1997)}]{Janssen97}%
  \BibitemOpen
  \bibfield  {author} {\bibinfo {author} {\bibfnamefont {H.~K.}\ \bibnamefont {Janssen}},\ }\bibfield  {title} {\bibinfo {title} {On critical exponents and the renormalization of the coupling constant in growth models with surface diffusion},\ }\href {https://doi.org/10.1103/PhysRevLett.78.1082} {\bibfield  {journal} {\bibinfo  {journal} {Phys. Rev. Lett.}\ }\textbf {\bibinfo {volume} {78}},\ \bibinfo {pages} {1082} (\bibinfo {year} {1997})}\BibitemShut {NoStop}%
\bibitem [{\citenamefont {Zhou}\ \emph {et~al.}(2015)\citenamefont {Zhou}, \citenamefont {Li}, \citenamefont {Zhu}, \citenamefont {Zuo},\ and\ \citenamefont {Yang}}]{Xuzhou15}%
  \BibitemOpen
  \bibfield  {author} {\bibinfo {author} {\bibfnamefont {Y.}~\bibnamefont {Zhou}}, \bibinfo {author} {\bibfnamefont {Y.}~\bibnamefont {Li}}, \bibinfo {author} {\bibfnamefont {H.}~\bibnamefont {Zhu}}, \bibinfo {author} {\bibfnamefont {X.}~\bibnamefont {Zuo}},\ and\ \bibinfo {author} {\bibfnamefont {J.}~\bibnamefont {Yang}},\ }\bibfield  {title} {\bibinfo {title} {The three-point sinuosity method for calculating the fractal dimension of machined surface profile},\ }\href {https://doi.org/10.1142/S0218348X15500164} {\bibfield  {journal} {\bibinfo  {journal} {Fractals}\ }\textbf {\bibinfo {volume} {23}},\ \bibinfo {pages} {1550016} (\bibinfo {year} {2015})}\BibitemShut {NoStop}%
\bibitem [{\citenamefont {Reis}(2015)}]{Reis2015}%
  \BibitemOpen
  \bibfield  {author} {\bibinfo {author} {\bibfnamefont {F.~D. A.~A.}\ \bibnamefont {Reis}},\ }\bibfield  {title} {\bibinfo {title} {Scaling of local roughness distributions},\ }\href {https://doi.org/10.1088/1742-5468/2015/11/P11020} {\bibfield  {journal} {\bibinfo  {journal} {Journal of Statistical Mechanics: Theory and Experiment}\ }\textbf {\bibinfo {volume} {2015}},\ \bibinfo {pages} {P11020} (\bibinfo {year} {2015})}\BibitemShut {NoStop}%
\end{thebibliography}

%apsrev4-2.bst 2019-01-14 (MD) hand-edited version of apsrev4-1.bst
%Control: key (0)
%Control: author (8) initials jnrlst
%Control: editor formatted (1) identically to author
%Control: production of article title (0) allowed
%Control: page (0) single
%Control: year (1) truncated
%Control: production of eprint (0) enabled
%

\end{document}